# Investigation of Hill Cipher Modifications Based on Permutation and Iteration

Mina Farmanbar
Dept. of Computer Engineering
Eastern Mediterranean University
Famagusta T.R. North Cyprus via Mersin 10, Turkey
Mina.farmanbar@emu.edu.tr

Alexander G. Chefranov
Dept. of Computer Engineering
Eastern Mediterranean University
Famagusta T.R. North Cyprus via Mersin 10, Turkey
Alexander.chefranov@emu.edu.tr

*Abstract*—Two recent Hill cipher modifications which iteratively use interweaving and interlacing are considered. We show that strength of these ciphers is due to non-linear transformation used in them (bit-level permutations). Impact of number of iterations on the avalanche effect is investigated. We propose two Hill cipher modifications using column swapping and arbitrary permutation with significantly less computational complexity (2 iterations are used versus 16). The proposed modifications decrease encryption time while keeping the strength of the ciphers. Numerical experiments for two proposed ciphers indicate that they can provide a substantial avalanche effect.

## I. INTRODUCTION

In the Hill cipher [1], ciphertext C is obtained by multiplication of a plaintext vector P by a key matrix, K, i.e., by a linear transformation. Encryption is given by:

$$C = KP(\bmod N), \qquad (1)$$

and decryption by:

$$P = K^{-1}C(\bmod N), \qquad (2)$$

where $K^{-1}$ is the modular arithmetic inverse of K, N>1. It can be broken by known plaintext-ciphertext attack due to its linearity. There are cryptosystems [2, 3, 4, 5, 6, 7] which have been developed in order to modify the Hill cipher to achieve higher security. In them, the Hill cipher is modified by including interweaving, interlacing, and iteration. They have significant avalanche effect and are supposed to resist cryptanalytic attacks. Strength of the ciphers is supposed to come from the nonlinearity of the m times applied matrix multiplication followed by interlacing or interweaving as it is mentioned explicitly or implicitly in [2, 3, 4, 5, 6, 8]. In [8] only, nonlinearity is related to the number of iterations m defining the order of the system of non-linear equations with respect to elements of the key matrix, the role of used permutations (interlacing, interweaving is not mentioned at all as a source of nonlinearity. If no permuation is used, also non-linear equations will be obtained for the key matrix elements after m iterations. However, resulting transformation is still linear, it may be represented by some matrix, and there is no need to solve non-linear equations to find elements of the original key matrix. For the cipher breaking, it is sufficient to define just the matrix resulting after several iterative multiplications. In all mentioned above papers, role of used permutations for non-linearity generation is not shown, and used in all the ciphers number of iterations m=16 is selected, we guess, on the base of discussion in [8]: "If we continue the process of iteration and take m=16, then we get 112 nonlinear equation of degree 16. As it is totally impossible to solve such a system of 112 non-linear equations, breaking the cipher is completely ruled out. Thus the cipher cannot be broken by the known plaintext attack." It is not discussed why interweaving and interlacing strengthen the Hill cipher.

In the present paper, we show that strength of the ciphers cipher modifications using interlacing, HCML [3], and interweaving, HCMW [5] is due to non-linear transformation used in it (bit-level permutations: interweaving and interlacing), investigate impact of number of iterations on the avalanche effect, and propose generalizations of the ciphers from [3, 5]. Then we present two new Hill cipher modifications which use bit-level permutations and only 1 or 2 iterations. We show that in the case of performing a bit-level permutation that swaps arbitrary selected bits, even two bits, a substantial avalanche effect is achieved.

The rest of the paper is organized as follows. First, a review of two Hill cipher modifications is given. Next, investigation of the number of iterations, experimental analysis and results of taking different number of iterations are presented. Then, two ciphers, column_swapping Hill cipher (CSHC) and arbitrary permutation Hill cipher (APHC) are proposed and their statistical analysis is conducted and discussed. Finally, we conclude the study. Appendix contains proof of non-linearity of bit-level permutations.

## II. REVIEW OF HILL CIPHER MODIFICATIONS

Hill cipher modifications HCML [3] and HCMW [5] use, respectively, interlacing and interweaving (transposition of the binary bits of the plaintext letters) and iteration. They are described as follows:

Input:

A plaintext of 2n 7-bit ASCII characters:

$$P = [P_{i,j}], i = 1 \text{ to } n, j = 1 \text{ to } 2 \quad (3)$$

and a key matrix K, such that each its entry is less than 64 used in HCML [3], and is less than 128 used in HCMW [5]:

$$K = [K_{i,j}], i = 1 \text{ to } n, j = 1 \text{ to } n. \quad (4)$$

HCML/HCMW encryption (N=128):

1. $P^0 = P$  (5)

2. For i = 1 to m where m=16 do the following:
   Compute, $P^i = KP^{i-1}$ (mod N)
   $P^i$ = interlace ($P^i$) as used in HCML [3], or $P^i$ = interweave ($P^i$) as used in HCMW [5].

3. $C = KP^m \bmod N$  (6)

Algorithm for interlace (P):

1. Divide P into two binary n×7 matrices, B and D, where $B_{k,j} = P_{k,j}$ and $D_{k,j} = P_{k,j+7}$, k = 1 to n, j = 1 to 7.
2. Mix $B_{k,j}$ and $D_{k,j}$ to get two binary n×7 matrices, $B'$ and $D'$, so that each $B_{k,j}$ lies in them adjacent to its corresponding $D_{k,j}$ as:

$$B' = \begin{bmatrix} b_{1,1} & d_{1,1} & \ldots & \ldots & b_{1,3} & d_{1,3} & b_{1,4} \\ d_{1,4} & \ldots & \ldots & \ldots & \ldots & b_{1,7} & d_{1,7} \\ \vdots & \vdots & \vdots & \vdots & \vdots & \vdots & \vdots \\ b_{n/2,1} & d_{n/2,1} & \ldots & \ldots & b_{n/2,3} & d_{n/2,3} & b_{n/2,4} \\ d_{n/2,4} & \ldots & \ldots & \ldots & \ldots & b_{n/2,7} & d_{n/2,7} \end{bmatrix}$$

$$D' = \begin{bmatrix} b_{\frac{n}{2}+1,1} & d_{\frac{n}{2}+1,1} & \ldots & \ldots & b_{\frac{n}{2}+1,3} & d_{\frac{n}{2}+1,3} & b_{\frac{n}{2}+1,4} \\ d_{\frac{n}{2}+1,4} & \ldots & \ldots & \ldots & \ldots & b_{\frac{n}{2}+1,7} & d_{\frac{n}{2}+1,7} \\ \vdots & \vdots & \vdots & \vdots & \vdots & \vdots & \vdots \\ b_{n,1} & d_{n,1} & \ldots & \ldots & b_{n,3} & d_{n,3} & b_{n,4} \\ d_{n,4} & \ldots & \ldots & \ldots & \ldots & b_{n,7} & d_{n,7} \end{bmatrix}$$

3. Construct $P_{j,1}$ from $B'_{j,1:7}$ and $P_{j,2}$ from $D'_{j,1:7}$ and convert them to decimal form, j = 1 to n

Algorithm for interweave (P):

1. Convert P into a binary n×14 matrix:

$$B = \begin{bmatrix} b_{1,1} & \cdots & b_{1,14} \\ \vdots & & \vdots \\ b_{n,1} & \cdots & b_{n,14} \end{bmatrix}$$

2. Rotate circular upward the jth column of B to get new column as $\begin{bmatrix} b_{2,j} \\ b_{3,j} \\ \vdots \\ b_{n,j} \\ b_{1,j} \end{bmatrix}$ where j = 1,3,5...

3. Similarly, rotate circular leftward the jth row of B where j = 2,4,6,....

4. Construct P from B using first 7 bits of jth row for $P_{j,1}$ and last 7 bits for $P_{j,2}$, j = 1,2,…,n

In the proposed algorithms, both interweaving and interlacing are the types of the bit-level permutation which makes total transformation non-linear that defines strength of these ciphers. A proof of non-linearity of a transformation represented by a bit-level permutation is given in Appendix 1.

Let's consider an example, in which a bit-level permutation is used after matrix multiplication showing that known plaintext-ciphertext attack is non-applicable even in the case of a trivial bit-level permutation that just swaps two bits.

We use in the example below m=26, a 2×2 key matrix $K = \begin{bmatrix} 19 & 12 \\ 21 & 13 \end{bmatrix}$, a pair of plaintext-ciphertext matrices $X_1 = \begin{bmatrix} 12 & 3 \\ 5 & 4 \end{bmatrix}$, $Y_1 = \begin{bmatrix} 2 & 1 \\ 5 & 11 \end{bmatrix}$, and $X_2 = \begin{bmatrix} 4 & 2 \\ 12 & 3 \end{bmatrix}$ which is considered as a new plaintext block.

We denote the permuted matrix as:

$Y'_i$ = bit_level permutation $(Y_i, P)$
where, $Y_i$ is a ciphertext matrix obtained for $X_i$, i = 1,2, P is a permutation.

**Example:**

Let $Y''_1 = \begin{bmatrix} 2 & 1 \\ 3 & 11 \end{bmatrix}$ be a result of a bit-level permutation swapping two bits, $b_2$ and $b_1$, of the $Y_{1i,j} = b_4 b_3 b_2 b_1 b_0$ where i = 2, j =1, i.e. the permutation is P=(43120) out of five bits. So the key can be obtained by an opponent after setting a linear system and solving it as $K_1 = \begin{bmatrix} 19 & 12 \\ 10 & 13 \end{bmatrix}$.

For $X_2$ as a new plaintext, $Y'_2 = \begin{bmatrix} 12 & 22 \\ 6 & 3 \end{bmatrix}$ is the permuted ciphertext. But $K_1^{-1}Y'_2$ mod N= $\begin{bmatrix} 24 & 12 \\ 2 & 11 \end{bmatrix}$ is not equal to $X_2$, where $K_1^{-1}$ mod N = $\begin{bmatrix} 13 & 4 \\ 12 & 11 \end{bmatrix}$.

III. INVESTIGATION OF NUMBER OF ITERATIONS IN HCML AND HCMW

In the HCML/HCMW, m=16 iterations are used to ensure the security and provide a good avalanche effect, i.e. changing one bit of the plaintext or one bit of the key should produce change in a lot of bits of the ciphertext. The number of iterations m is taken to be 16 [8] because of having in that case non-linear system of equations of 16-th order, but actaully it is not the source of non-linearity of the used transformations. Non-linearity of the transformations used in the ciphers under consideration comes from the use of bit-level permutations (their non-linearity is proved in Appendix). Hence, may be with less number of iterations, still avalanche effect is good.

We examine avalanche effect of these ciphers using examples of plaintext and key from [3, 5] for different number of iterations.

Plaintext, given by (7):

"The World Bank h" (7)

and key by (8), are from [3]:

$$K1 = \begin{bmatrix} 53 & 62 & 24 & 33 & 49 & 18 & 17 & 43 \\ 45 & 12 & 63 & 29 & 60 & 35 & 58 & 11 \\ 8 & 41 & 46 & 30 & 48 & 32 & 5 & 51 \\ 47 & 9 & 38 & 42 & 2 & 59 & 27 & 61 \\ 57 & 20 & 6 & 31 & 16 & 26 & 22 & 25 \\ 56 & 37 & 13 & 52 & 3 & 54 & 15 & 21 \\ 36 & 40 & 44 & 10 & 19 & 39 & 55 & 4 \\ 14 & 1 & 23 & 50 & 34 & 0 & 7 & 28 \end{bmatrix} \quad (8)$$

and plaintext (9):

"The development", (9)

and key (10) are from [5],

$$K2 = \begin{bmatrix} 53 & 62 & 124 & 33 & 49 & 118 & 107 & 43 \\ 45 & 112 & 63 & 29 & 60 & 35 & 58 & 11 \\ 88 & 41 & 46 & 30 & 48 & 32 & 105 & 51 \\ 47 & 99 & 38 & 42 & 112 & 59 & 27 & 61 \\ 57 & 20 & 6 & 31 & 106 & 126 & 22 & 125 \\ 56 & 37 & 113 & 52 & 3 & 54 & 105 & 21 \\ 36 & 40 & 43 & 100 & 119 & 39 & 55 & 94 \\ 14 & 81 & 23 & 50 & 34 & 70 & 7 & 28 \end{bmatrix} \quad (10)$$

There are some problems in the example from [5] illustrating the avalanche effect. The plaintext (9) in ASCII code shall have letter "l" represented by 108 that in [5] is shown as 109. Correct ASCII code representation for (9) is given in (11):

$$P = \begin{bmatrix} 84 & 108 \\ 104 & 111 \\ 101 & 112 \\ 32 & 109 \\ 100 & 101 \\ 101 & 110 \\ 118 & 116 \\ 101 & 32 \end{bmatrix} \quad (11)$$

Correct result after multiplication taking into account (11) is given by:

$$P^1 = \begin{bmatrix} 27 & 112 \\ 17 & 83 \\ 83 & 113 \\ 34 & 73 \\ 37 & 25 \\ 38 & 86 \\ 86 & 77 \\ 127 & 11 \end{bmatrix} \quad (12)$$

Table 1 shows comparison results that were obtained by changing the first character of the plaintext (7) from "T" to "U" and the 9th character of the plaintext (10) from "l" to "m" for different number of iterations ranging from 1 to 100. We also change the key (8) element $K1_{3,3}$ from 46 to 47 and the key (9) element $K2_{3,6}$ from 32 to 33.

From Table 1, we can see that for all number of iterations avalanche effect is approximately the same. Hence, used in HCML/HCML number of iterations equal to 16 is not distinguished and less number of iterations may be used instead.

TABLE 1. AVALANCHE EFFECT INVESTIGATION FOR HCML AND HCMW

| m | Change in plaintext | | Change in key | |
|---|---|---|---|---|
| | *Number of bits that differ* | | *Number of bits that differ* | |
| | HCML | HCMW | HCML | HCMW |
| 1 | 56 | 64 | 30 | 51 |
| 2 | 52 | 59 | 55 | 61 |
| 3 | 53 | 54 | 57 | 59 |
| 4 | 56 | 53 | 58 | 55 |
| 5 | 53 | 40 | 56 | 56 |
| 6 | 62 | 61 | 58 | 56 |
| 7 | 57 | 59 | 59 | 48 |
| 8 | 61 | 54 | 62 | 61 |
| 9 | 44 | 63 | 61 | 62 |
| 10 | 62 | 62 | 47 | 60 |
| 11 | 53 | 64 | 51 | 54 |
| 12 | 56 | 60 | 60 | 56 |
| 13 | 57 | 50 | 49 | 66 |
| 14 | 52 | 54 | 57 | 64 |
| 15 | 60 | 62 | 61 | 57 |
| 16 | 65 | 43 | 55 | 57 |
| 17 | 51 | 60 | 66 | 56 |
| 18 | 51 | 60 | 53 | 62 |
| 19 | 68 | 53 | 62 | 50 |
| 20 | 59 | 59 | 57 | 53 |
| 50 | 58 | 63 | 56 | 49 |
| 100 | 59 | 53 | 58 | 61 |

## IV. PROPOSED CIPHERS

We introduce Column_swapping Hill cipher (CSHC). It uses swapping columns of the binary bits of the plaintext characters instead of interlacing and interweaving as in [3, 5]. Also, we introduce arbitrary permutation Hill cipher (APHC) that uses an arbitrary permutation not known to an opponent and shared between the two communication parties instead of a fixed permutation (interweaving or interlacing). In CSHC and APHC, 1 or 2 iterations are used instead of 16 iterations used in [3, 5] Cipher inputs are the same as used in HCML/HCMW, but there are some additional inputs:

- Number of iterations m is considerd as m∈{1,2} instead of 16

- Permutation that is a vector of the same length as P (i.e., L = n×14) with integer components from {1,...,L}. All values from 1,...L are represented in Permutation in some order. For example, if L=4 and Permutation=(4,1,3,2) then applying Permutation to P=$(b_0, b_1, b_2, b_3)$, we get $(b_3, b_0, b_2, b_1)$.
- Additional_multiplication (AD) which has two values True/False and defines whether the last multiplication in the algorithms is to be applied.

Algorithm for Column_swapping (P):

1. Divide P into two binary n×7 matrices, E and F, where $E_{k,j} = P_{k,j}$ and $F_{k,j} = P_{k,j+7}$, k = 1 to n, j = 1 to 7.

$$E = \begin{bmatrix} e_{1,1} & \cdots & e_{1,7} \\ \vdots & & \vdots \\ e_{n,1} & \cdots & e_{n,7} \end{bmatrix}, F = \begin{bmatrix} f_{1,1} & \cdots & f_{1,7} \\ \vdots & & \vdots \\ f_{n,1} & \cdots & f_{n,7} \end{bmatrix}$$

2. Swap the j-th column of E and jth column of F where j = 2,4,6 as shown below for n=8:

$$E' = \begin{bmatrix} e_{11} & f_{12} & e_{13} & f_{14} & e_{15} & f_{16} & e_{17} \\ e_{21} & f_{22} & e_{23} & f_{24} & e_{25} & f_{26} & e_{27} \\ e_{31} & f_{32} & e_{33} & f_{34} & e_{35} & f_{36} & e_{37} \\ e_{41} & f_{42} & e_{43} & f_{44} & e_{45} & f_{46} & e_{47} \\ e_{51} & f_{52} & e_{53} & f_{54} & e_{55} & f_{56} & e_{57} \\ e_{61} & f_{62} & e_{63} & f_{64} & e_{65} & f_{66} & e_{67} \\ e_{71} & f_{72} & e_{73} & f_{74} & e_{75} & f_{76} & e_{77} \\ e_{81} & f_{82} & e_{83} & f_{84} & e_{85} & f_{86} & e_{87} \end{bmatrix}$$

$$F' = \begin{bmatrix} f_{11} & e_{12} & f_{13} & e_{14} & f_{15} & e_{16} & f_{17} \\ f_{21} & e_{22} & f_{23} & e_{24} & f_{25} & e_{26} & f_{27} \\ f_{31} & e_{32} & f_{33} & e_{34} & f_{35} & e_{36} & f_{37} \\ f_{41} & e_{42} & f_{43} & e_{44} & f_{45} & e_{46} & f_{47} \\ f_{51} & e_{52} & f_{53} & e_{54} & f_{55} & e_{56} & f_{57} \\ f_{61} & e_{62} & f_{63} & e_{64} & f_{65} & e_{66} & f_{67} \\ f_{71} & e_{72} & f_{73} & e_{74} & f_{75} & e_{76} & f_{77} \\ f_{81} & e_{82} & f_{83} & e_{84} & f_{85} & e_{86} & f_{87} \end{bmatrix}$$

3. Set $P_{j,1} = E'_{j,1:7}$ and $P_{j,2} = E'_{j,1:7}$ where j = 1 to n

Algorithm for APHC (Permutation, P):

1. Convert P into a binary n×14 matrix:

$$B = \begin{bmatrix} b_{1,1} & \cdots & b_{1,14} \\ \vdots & & \vdots \\ b_{n,1} & \cdots & b_{n,14} \end{bmatrix}$$

2. Apply Permutation to the bits of B that is considered as a row-vector = (v1,v2,...vn×14) obtained in row-major order.
3. Construct P from B using first 7 bits of j-th row for $P_{j,1}$ and last 7 bits for $P_{j,2}$ where j = 1 to n.

The CSHC and APHC ciphers are shown as a diagram in Fig. 1.

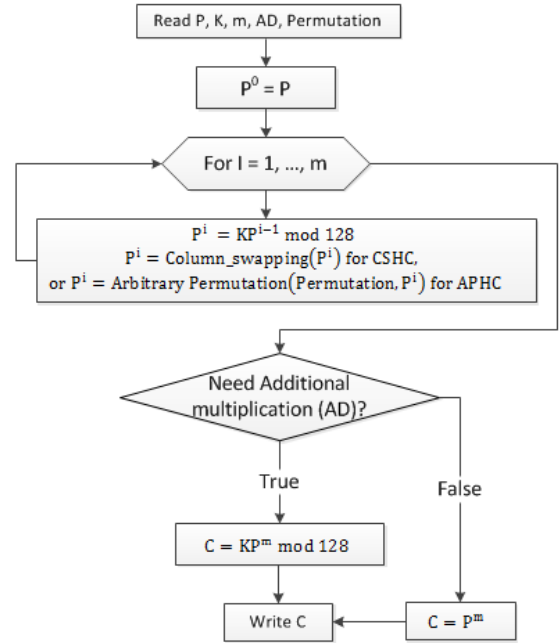

Figure 1. Schematic diagram of the CSHC and APHC. Here, m denotes the number of iterations, and m∈{1,2}.

For the proposed ciphers, in the case of CSHC with AD=False and m=1, ciphertext C is defined as follows C=Column_swapping(K*P).
If an opponent applies to C inverse of Column_swapping permutation, he gets K*P, and, hence, the key K of the algorithm can be disclosed by the opponent by the known plaintext-ciphertext attack. In the case of AD=True or m=2, such attack is not possible. In the case of APHC, iteration number may be taken m=1 with AD=False since a permutation applied in it is kept secret, and thus, can not be inverted without enumeration of possible permutations number of which exponentially grows with the size L of the permuted vector. Hence, key space for APHC is L! times greater than that of CSHC and HC.

Let us illustrate the CSHC algorithm after multiplying plaintext (9) and the key (10) and getting (12). After dividing (12) into two binary matrices, we get:

$$E = \begin{bmatrix} 0 & 0 & 1 & 1 & 0 & 1 & 1 \\ 0 & 0 & 1 & 0 & 0 & 0 & 1 \\ 1 & 0 & 1 & 0 & 0 & 1 & 1 \\ 0 & 1 & 0 & 0 & 0 & 1 & 0 \\ 0 & 1 & 0 & 0 & 1 & 0 & 1 \\ 0 & 1 & 0 & 0 & 1 & 1 & 0 \\ 1 & 0 & 1 & 0 & 1 & 1 & 0 \\ 1 & 1 & 1 & 1 & 1 & 1 & 1 \end{bmatrix}$$

$$F = \begin{bmatrix} 1 & 1 & 1 & 0 & 0 & 0 & 0 \\ 1 & 0 & 1 & 0 & 0 & 1 & 1 \\ 1 & 1 & 1 & 0 & 0 & 0 & 1 \\ 1 & 0 & 0 & 1 & 0 & 0 & 1 \\ 0 & 0 & 1 & 1 & 0 & 0 & 1 \\ 1 & 0 & 1 & 0 & 1 & 1 & 0 \\ 1 & 0 & 0 & 1 & 1 & 0 & 1 \\ 0 & 0 & 0 & 1 & 0 & 1 & 1 \end{bmatrix}$$

Now, we show the process of CSHC:

$$E' = \begin{bmatrix} 0 & 1 & 1 & 0 & 0 & 0 & 1 \\ 0 & 0 & 1 & 0 & 0 & 1 & 1 \\ 1 & 1 & 1 & 0 & 0 & 0 & 1 \\ 0 & 0 & 0 & 1 & 0 & 0 & 0 \\ 0 & 0 & 0 & 1 & 1 & 0 & 1 \\ 0 & 0 & 0 & 0 & 1 & 1 & 0 \\ 1 & 0 & 1 & 1 & 1 & 0 & 0 \\ 1 & 0 & 1 & 1 & 1 & 1 & 1 \end{bmatrix}$$

$$F' = \begin{bmatrix} 1 & 0 & 1 & 1 & 0 & 1 & 0 \\ 1 & 0 & 1 & 0 & 0 & 0 & 1 \\ 1 & 0 & 1 & 0 & 0 & 1 & 1 \\ 1 & 1 & 0 & 0 & 0 & 1 & 1 \\ 0 & 1 & 1 & 0 & 0 & 0 & 1 \\ 1 & 1 & 1 & 0 & 1 & 1 & 0 \\ 1 & 0 & 0 & 0 & 1 & 1 & 1 \\ 0 & 1 & 0 & 1 & 0 & 1 & 1 \end{bmatrix}$$

Transformed plaintext, after the first iteration is as follows:

$$P = \begin{bmatrix} 49 & 90 \\ 19 & 81 \\ 113 & 83 \\ 8 & 99 \\ 13 & 49 \\ 6 & 118 \\ 92 & 71 \\ 95 & 43 \end{bmatrix}$$

To illustrate APHC let $(b_6, b_5, b_4, b_3, b_0, b_2, b_1)$ be a result of 3-bit permutation by swapping three bits $b_2, b_1$ and $b_0$ out of the 7-bit ASCII code binary represented by $b_6 b_5 b_4 b_3 b_2 b_1 b_0$. After converting (12) into a binary matrix, we get:

$$E = \begin{bmatrix} 0 & 0 & 1 & 1 & 0 & 1 & 1 & 1 & 1 & 0 & 0 & 0 & 0 \\ 0 & 0 & 1 & 0 & 0 & 0 & 1 & 1 & 0 & 1 & 0 & 0 & 1 & 1 \\ 1 & 0 & 1 & 0 & 0 & 1 & 1 & 1 & 1 & 0 & 0 & 0 & 1 \\ 0 & 1 & 0 & 0 & 0 & 1 & 0 & 1 & 0 & 0 & 1 & 0 & 0 & 1 \\ 0 & 1 & 0 & 0 & 1 & 0 & 1 & 0 & 0 & 1 & 1 & 0 & 0 & 1 \\ 0 & 1 & 0 & 0 & 1 & 1 & 0 & 1 & 0 & 1 & 0 & 1 & 1 & 0 \\ 1 & 0 & 1 & 0 & 1 & 1 & 0 & 1 & 0 & 0 & 1 & 1 & 0 & 1 \\ 1 & 1 & 1 & 1 & 1 & 1 & 0 & 0 & 0 & 1 & 0 & 1 & 1 \end{bmatrix}$$

The process of APHC after performing P= (6,5,4,3,0,2,1) on the $e_{i,j}$ where i = 1, j = 1 to 7:

$$E' = \begin{bmatrix} 0 & 0 & 1 & 1 & 1 & 0 & 1 & 1 & 1 & 1 & 0 & 0 & 0 & 0 \\ 0 & 0 & 1 & 0 & 0 & 0 & 1 & 1 & 0 & 1 & 0 & 0 & 1 & 1 \\ 1 & 0 & 1 & 0 & 0 & 1 & 1 & 1 & 1 & 0 & 0 & 0 & 1 \\ 0 & 1 & 0 & 0 & 0 & 1 & 0 & 1 & 0 & 0 & 1 & 0 & 0 & 1 \\ 0 & 1 & 0 & 0 & 1 & 0 & 1 & 0 & 0 & 1 & 1 & 0 & 0 & 1 \\ 0 & 1 & 0 & 0 & 1 & 1 & 0 & 1 & 0 & 1 & 0 & 1 & 1 & 0 \\ 1 & 0 & 1 & 0 & 1 & 1 & 0 & 1 & 0 & 0 & 1 & 1 & 0 & 1 \\ 1 & 1 & 1 & 1 & 1 & 1 & 0 & 0 & 0 & 1 & 0 & 1 & 1 \end{bmatrix}$$

plaintext matrix, after the first iteration is as follows:

$$P^1 = \begin{bmatrix} 29 & 112 \\ 17 & 83 \\ 83 & 113 \\ 108 & 41 \\ 37 & 25 \\ 38 & 86 \\ 59 & 61 \\ 127 & 11 \end{bmatrix}$$

V. STATISTICAL ANALYSIS OF THE PROPOSED CSHC AND APHC

To test the strength of the CSHC and APHC we examine both changing elements in the plaintext and key. Table 2 shows avalanche effect of CSHC when changing first character of the plaintext (9) from "T" to "U" which differ by one bit, then changing second character from "h" to "i" and so on where m∈{1,2} and additional multiplication AD is true. We also change the key (10) element $K2_{3,6}$ from 32 to 33.

From Table 2, we can see for CSHC that after m iterations avalanche effect is more or less the same where m∈{1,2}. Hence, one iteration can be sufficient i.e., m = 1.

Table 3 shows the avalanche effect average of 17 samples for APHC that swaps selected z bits of both plaintext (9) by changing "T" to "U" and key (10) by changing $K2_{3,6}$ element from 32 to 33 by performing iteration and additional multiplication AD i.e., m∈{1,2} and AD = True/False to determine how changing bits provides avalanche effect where z = 2 to 7.

TABLE 2. AVALANCHE EFFECT OF CSHC WHERE ADDITIONAL MULTIPLICATION AD = TRUE AND M∈{1,2}

| Plaintext characters | Original key | | Changed key | |
|---|---|---|---|---|
| | m =1 AD = true | m = 2 AD = true | m =1 AD = true | m = 2 AD = true |
| "T" to "U" | 44 | 44 | 46 | 64 |
| "h" to "i" | 42 | 55 | 60 | 61 |
| "e" to "f" | 40 | 55 | 59 | 49 |
| "d" to "e" | 60 | 60 | 61 | 66 |
| "e" to "f" | 56 | 45 | 51 | 61 |
| "v" to "w" | 55 | 50 | 55 | 52 |
| "e" to "f" | 56 | 45 | 49 | 49 |
| "l" to "m" | 51 | 51 | 56 | 62 |
| "o" to "p" | 48 | 51 | 56 | 56 |
| "p" to "q" | 49 | 47 | 64 | 65 |
| "m" to "n" | 51 | 58 | 53 | 57 |
| "e" to "f" | 47 | 54 | 60 | 57 |
| "n" to "o" | 53 | 44 | 65 | 55 |
| "t" to "u" | 45 | 42 | 63 | 50 |

TABLE 3. AVALANCHE EFFECT AVERAGE OF APHC WHERE ADDITIONAL MULTIPLICATION AD = TRUE/ FALSE AND M ∈ {1,2}

| z | Change in plaintext | | | Change in key | | |
|---|---|---|---|---|---|---|
|   | m = 1 AD = false | m = 1 AD = true | m = 2 AD = false | m = 1 AD = true | m = 1 AD = false | m = 2 AD = true |
| 2 | 36.6 | 45.1 | 50.1 | 9.41 | 57.5 | 58.8 |
| 3 | 35.8 | 45.5 | 50.8 | 10.6 | 56.7 | 60.1 |
| 4 | 36.0 | 39.8 | 49.8 | 11.3 | 55.2 | 63.1 |
| 5 | 37.1 | 38.0 | 51.1 | 11.0 | 55.1 | 63.1 |
| 6 | 36.8 | 43.0 | 48.3 | 10.7 | 60.5 | 55.5 |
| 7 | 36.6 | 44.2 | 49.2 | 14.8 | 55.2 | 56.5 |

TABLE 4. PERMUTATION ORDERS AND BIT LOCATIONS

| z | Permutation | Element indices |
|---|---|---|
| 2 | 6453210 | $P_{11}, P_{31}$ |
| 3 | 6543021 | $P_{32}, P_{71}$ |
| 4 | 6234510 | $P_{72}, P_{12}$ |
| 5 | 2345160 | $P_{72}, P_{52}$ |
| 6 | 1234560 | $P_{62}, P_{41}$ |
| 7 | 0123456 | $P_{32}, P_{12}$ |

Table 4 displays the number of swapped bits z and Permutations which were applied when getting the avalanche effect average of used samples in the Table 3 on the plaintext characters P that are represented as 7-bit binary $(b_6 b_5 b_4 b_3 b_2 b_1 b_0)$. For example two bits $b_5, b_4$ are swapped in Permutation(6,4,5,3,2,1,0) that is applied on both elements $P_{ij}$ in the plaintext matrix where i=1,3 and j=1.

We have seen that even a small change in the plaintext or key results in changing approximately half of the ciphertext bits. From Table 3, we found that any simple bit-level permutation can provide a substantial avalanche effect same as other complicated and fixed permutations which have been used in the HCML and HCMW.

## VI. CONCLUSION

The Hill cipher is susceptible to known plaintext-ciphertext attack due to its linearity. In this study, we generalized two Hill cipher modifications [3, 5] which use bit-level permutation and 16 iterations. In both cases, the Hill cipher has been made secure against the attack. We proved that strength of the ciphers is due to non-linear transformation used in them (bit-level permutations), and we found that, for number of iterations from 1 to 100, avalanche effect is approximately the same. Hence, use of 16 iterations is not reasonable, and less number of iterations may be used instead. We proposed two new Hill cipher modifications, CSHC and APHC, that also use bit-level permutation and one or two iterations. Results of statistical tests for examining the strength of CSHC and APHC are given which indicate that any bit-level permutation can provide a substantial avalanche effect.

## APPENDIX

Here we show the non-linearity of the bit-level transposition $P$ swapping $i$-th and $j$-th bits in a binary number

$$b = \sum_{l=0}^{n} b_l 2^l$$

$b' = P(b),$
$b = (b_n b_{n-1} .. b_{i+1} b_i b_{i-1} .. b_{j+1} b_j b_{j-1} .. b_1 b_0),$
$b' = (b_n b_{n-1} .. b_{i+1} b_j b_{i-1} .. b_{j+1} b_i b_{j-1} .. b_1 b_0)$

A linear transformation satisfies the following:

$$T(a_1 X + a_2 Y) = a_1 T(X) + a_2 T(Y) , \quad (13)$$

where, $a_1$, $a_2$ are any scalars, and $X, Y$ are any two objects to which transformation $T$ is applicable. Let us show that the binary permutation P does not meet (13) for $a_1 = a_2 = 1$ and some two binary numbers,

$b^1 = (b_n^1 .. b_{i+1}^1 b_i^1 b_{i-1}^1 .. b_{j+1}^1 b_j^1 b_{j-1}^1 .. b_0^1),$
$b^2 = (b_n^2 .. b_{i+1}^2 b_i^2 b_{i-1}^2 .. b_{j+1}^2 b_j^2 b_{j-1}^2 .. b_0^2)$

where these numbers are selected so that $b_i^l = 0, b_j^l = 1, b_{j-1}^l = 0, l = \overline{1,2}, b_l^1 = 0, b_l^2 = 1, l = \overline{j+1, i-1},$
Then,
$P(b^1 + b^2) = P((b_n^1 .. b_{i+1}^1 00..010 b_{j-2}^1 .. b_0^1) +$
$(b_n^2 .. b_{i+1}^2 01..110 b_{j-2}^2 .. b_0^2)) =$
$P(b^3) = P(b_n^3 .. b_{i+1}^3 11..10 b_{j-1}^3 .. b_0^3) =$
$(b_n^3 .. b_{i+1}^3 01..11 b_{j-1}^3 .. b_0^3)$

From the other side,
$P(b^1) + P(b^2) = b^4 + b^5 = (b_n^1 .. b_{i+1}^1 b_j^1 b_{i-1}^1 .. b_{j+1}^1 b_i^1 b_{j-1}^1 .. b_0^1)$
$+ (b_n^2 .. b_{i+1}^2 b_j^2 b_{i-1}^2 .. b_{j+1}^2 b_i^2 b_{j-1}^2 .. b_0^2) =$
$(b_n^1 .. b_{i+1}^1 10..000 b_{j-2}^1 .. b_0^1) + (b_n^2 .. b_{i+1}^2 11..100 b_{j-2}^2 .. b_0^2) =$
$P(b_n^6 .. b_{i+1}^6 01..10 b_{j-1}^3 .. b_0^3) = (b_n^6 .. b_{i+1}^6 01..10 b_{j-1}^3 .. b_0^3) =$
$b^6 \neq P(b^3)$

The last inequality proves that the transposition $P(b)$ swapping $i$-th and $j$-th bits in the binary representation of the number $b$ is a non-linear transformation, because for any transpostion we can construct two binary numbers such that (13) is violated for them and the transposition.
For example, let

$n = 4, i = 3, j = 2, b^1 = (10101) = 21,$

$b^2 = (00101) = 5, b^3 = (21+5) \mod 32 = (11010)$

$= 26,$

$P(b^3) = (10110) = 22,$

$P(b^1) = (11001) = 25, P(b^2) = (01001) = 9,$

$P(b^1) + P(b^2) = (11001) + (01001)$

$= 25 + 9 = 34 \mod 32 = 2 = (00010) = b^6 \neq P(b^3),$

As far as any permutation can be represented as a product of transpositions (see, e.g., http://en.wikipedia.org/wiki/Transposition_(mathematics)#Transpositions), we have proved that any binary-level permutation is a non-linear transformation.